\documentclass[prd,superscriptaddress,showpacs,floatfix,%
preprintnumbers,nofootinbib]{revtex4}
\usepackage{bm}
\usepackage{epsfig}
\usepackage[small]{caption}
\usepackage{graphicx}
\usepackage{amsmath}
\newcommand{\lwig}{\mbox{\;\raisebox{.3ex}
    {$<$}$\!\!\!\!\!$\raisebox{-.9ex}{$\sim$}\;}}

\def\beq{\begin{equation}}
\def\eeq{\end{equation}}
\def\bea{\begin{eqnarray}}
\def\eea{\end{eqnarray}}
\def\bi{\begin{itemize}}
\def\ei{\end{itemize}}

\def\bar#1{\overline{#1}}

\def\inv{^{\raise.15ex\hbox{${\scriptscriptstyle -}$}\kern-.05em 1}}
\def\lbar{{\lower.35ex\hbox{$\mathchar'26$}\mkern-10mu\lambda}} 
\def\e#1{{\rm e}^{^{\textstyle#1}}}

\def\to{\rightarrow}

\let\p=\partial
\let\<=\langle
\let\>=\rangle

\let\+=\uparrow

\long\def\dump#1{}

\let\ga=\gamma

\let\de=\delta

\let\th=\theta

\let\ta=\tau

\def\ph{\varphi}

\begin{document}


\title{Decaying neutrinos: The long way to isotropy}

\author{Anders Basb{\o}ll}
\email{ab421@sussex.ac.uk}
\affiliation{Department of Physics and Astronomy, University of Sussex, Brighton, BN1 9QH, United Kingdom}
\affiliation{Department of Physics and Astronomy, Aarhus University, DK-8000 Aarhus C, Denmark}
\author{Ole Eggers Bj{\ae}lde}
\email{bjaelde@physik.rwth-aachen.de}
\affiliation{Institut f\"ur Theoretische Physik E, RWTH Aachen University, D-52056 Aachen, Germany}
\affiliation{Department of Physics and Astronomy, Aarhus University, DK-8000 Aarhus C, Denmark}

\preprint{TTK-10-21}

\date{\today}

\begin{abstract}
We investigate a scenario in which neutrinos are coupled to a
pseudoscalar degree of freedom $\ph$ and where decays  $\nu_1 \to
\nu_2+\ph$ and inverse decays are the responsible mechanism for
obtaining equilibrium. In this context we discuss the implication
of the invisible neutrino decay on the neutrino-pseudoscalar coupling
constant and the neutrino lifetime. Assuming the realistic scenario of a
thermal background of neutrinos and pseudoscalar we update the bound 
on the (off-diagonal) neutrino-pseudoscalar coupling constant to $g<2.6\times10^{-13}$
and the bound on the neutrino lifetime to $\tau<1\times10^{13}\,\rm{s}$.
Furthermore we confirm analytically that kinetic equilibrium is delayed by two Lorentz
$\ga$--factors -- one for time dilation of the (decaying) neutrino
lifetime and one from the opening angle. We have also confirmed this behavior numerically.

\end{abstract}


\maketitle

\section{Introduction}

The possibility for neutrino interactions beyond the standard model
has been studied in many contexts over the years. One particularly
simple possibility is that neutrinos couple to a new pseudoscalar
degree of freedom, as is, for example, the case in Majoron models -
see the following references for previous discussions about the
dynamics of the strong neutrino-pseudoscalar coupling and its astrophysical
implications \cite{Chikashige:1980ui,Gelmini:1980re,Schechter:1981cv,Gelmini:1982rr,Kolb:1985tqa,Manohar:1987ec,Dicus:1988jh,Konoplich:1988,Berezhiani:1989za,Gelmini:1994az}.

Astrophysics provides fairly stringent constraints on such
couplings. For example SN1987A provides a bound on the
dimensionless coupling constant of order $10^{-7}\lwig g \lwig 10^{-5}$ 
\cite{Choi:1989hi,Raffelt:1996wa,Farzan:2002wx,Kachelriess:2000qc}
by requiring that the neutrino signal should not be significantly
shortened.

In the same way there are two cosmological bounds on $g$. First, the
value of $g$ should not be large enough that pseudoscalars are fully
thermalized before big bang nucleosynthesis. This leads to $g \lwig 10^{-5}$. Second, a
significant value of $g$ will make neutrinos self-interacting in the
late universe and prevent neutrino free-streaming. This possibility
has been discussed a number of times in the literature (see
\cite{Raffelt:1987ah,Beacom:2004yd,Hannestad:2004qu,Bell:2005dr,Sawyer:2006ju}).

The effect on cosmological observables such as the cosmic microwave background (CMB) spectrum were
studied in \cite{Beacom:2004yd,Hannestad:2004qu,Bell:2005dr,Sawyer:2006ju,Cirelli:2006kt,%
Friedland:2007vv,Hannestad:2005ex,Basboll:2008fx}); particularly it was found that
although models with no neutrino free-streaming can mimic the matter
power spectrum of $\Lambda$CDM models, they produce a distinct
signature in the CMB spectrum which is much harder to reproduce. The
feature arises because neutrinos act as a source term for photon
perturbations. If there is no free-streaming, the source term is
stronger and consequently the CMB anisotropy is increased for all
scales inside the particle horizon at recombination. On the other
hand, there is no effect on larger scales.

This distinct signature has been used to constrain models without
neutrino free-streaming and in
\cite{Hannestad:2005ex,Basboll:2008fx} it was used to constrain the
corresponding neutrino-pseudoscalar coupling parameters.

However, for decays and inverse decays the interaction was treated
in a somewhat simplified manner in the sense that the momentum
equilibration rate was assumed to be roughly $\Gamma^* \sim 1/(\ga^2
\ta)$, where $\ta$ is the rest-frame lifetime and $\gamma \sim
E_\nu/m_\nu$ is the Lorentz boost factor.

In this paper we wish to check this assumption in an explicit way with
a realistic setup. The possible departure from the simple relation
$\Gamma^* \sim 1/(\ga^2\ta)$ is something which is highly relevant for
parameter estimations - such as placing bounds on the neutrino-pseudoscalar
coupling and hence also for constraining the neutrino lifetime. Furthermore
it is something which needs to be taken into account in numerical studies
in which we allow for nonstandard neutrino interactions. One particular area
where detailed knowledge of the interaction would be useful is the search
for the cosmic neutrino background \cite{Hannestad:2009xu,de Bernardis:2007bu,Ringwald:2009bg,DeBernardis:2008ys}.

One comment is in order here: The motivation for looking at decays and inverse
decays rather than various scattering processes involving neutrinos
and pseudoscalars ($\nu\nu\to\ph\ph$,$\,\,\ph\ph\to\nu\nu$,$\,\,\nu\ph\to\nu\ph$) is that the probabilities
of these scattering processes are proportional to $g^4$, where $g$ is the
neutrino-pseudoscalar coupling constant. The probability of the decay $\nu_1 \to
\nu_2+\ph$, on the other hand, is proportional to $g^2$. Consequently, at small values of $g$
the decay actually dominates over the scattering processes and allows us to put severe constraints on $g$.

This paper is organized as follows: In Sec.~\ref{setup} we look
at the setup with a gas consisting of two neutrino species and
a pseudoscalar -- a gas that only has decays and inverse decays to obtain
equilibrium. We argue for the $\frac{1}{\tau\gamma^2}$ in the decay
rate. In Sec.~\ref{nobackg} we look at an initial situation of a
standing wave of the heavy neutrinos and no other particles. In
Sec.~\ref{thermback} we introduce thermal distributions of the
light neutrino and of pseudoscalars into a thermal background while keeping the
initial conditions for the heavy neutrino. Furthermore we discuss the
implication of the decay on the neutrino-pseudoscalar coupling and on the neutrino lifetime.
We present numerical results in Sec.~\ref{numerics}. Finally we have a conclusion and an appendix concerning
calculations for the numerical implementation of the system.

\section{Thermalization of a gas with only decays and inverse decays}\label{setup}

Thermalization of a gas by decay and inverse decay is a nontrivial
process because of phase space limitation. As long as one of the
involved particles can interact with an external heat bath, it is in
principle possible to thermalise the gas provided that the
interaction rate is sufficiently fast. This is, for instance, the case
with thermal leptogenesis in which the decay products are
thermalized by SM gauge interactions.

However, for the case studied here this is not true. The weak
interactions are far too weak to maintain equilibrium at the eV
temperatures considered here. In this case full thermal equilibrium
can never be achieved.

The standard case usually studied, for example, in the case of
thermal leptogenesis is a spatially homogeneous gas in which
interactions drive the distribution toward thermal equilibrium
(see e.g.\ \cite{Starkman:1993ik,Basboll:2006yx})

However, from the point of view of structure formation and more
specifically free-streaming the important point is the rate of
directional momentum transfer between species. For example, Thompson
scattering is inefficient for maintaining energy equilibration
between electrons and photons, but very efficient for exchanging
momentum between the two species. This can be seen from the simple
relations $|\Delta E_\gamma/E_\gamma| \sim E_\gamma/m_e$ and
$|\Delta \vec{p}/p| \sim 1$ in a single scattering event. Therefore
Thompson scattering is very efficient for driving the acoustic
baryon-photon oscillations prior to recombination.

However, for a gas with only decays and inverse decays momentum
transfer is even more inefficient than energy transfer. Roughly the
energy transfer time scale is given by the decay rate $\Gamma =
1/(\gamma \tau)$, i.e.\ the usual Lorentz suppressed rest-frame
decay rate. However, in the lab frame the decay products are emitted
in a cone of opening angle $1/\gamma$ relative to the direction of
momentum of the parent particle. Therefore, in a single decay the
momentum direction is changed by only $|\Delta \vec{p}/p| \sim
1/\gamma$. This finally means that the rate of momentum change in
the gas is roughly $1/(\gamma^2 \tau)$; i.e.\ for relativistic
decays it is highly suppressed, and even suppressed relative to the
energy exchange rate.

Let us begin with the Lagrangian for a generic pseudoscalar
neutrino interaction \footnote{The coupling structure could
in principle be derivative instead of pseudoscalar. However, this
point makes no difference to the discussion here since we study only
decays and inverse decays. One could also choose a scalar coupling -
it would only lead to a very small difference, which, in fact, is removed
completely in the approximation where the lighter neutrino is massless.}

 \beq \mathcal{L}=-i\sum_{j,k}g_{jk}\ph
\bar{\nu}_j \gamma_5 \nu_k. \eeq

We will consider only two neutrinos, one we consider to be heavy
($\nu_1$ [or just $1$ for convenience] with mass $m_1$), a massless
neutrino ($\nu_2$ [or just $2$ for convenience]) and a massless
pseudoscalar ($\ph$). Thus we drop the index of $g$ ($g\equiv g_{1,2}$)
and the Lagrangian becomes \beq \mathcal{L}=-ig\ph(
\bar{\nu}_1\ga_5\nu_2+\bar{\nu}_2\ga_5\nu_1). \eeq

In the following section we derive the specific Boltzmann collision
integrals relevant for decays and inverse decays in an inhomogeneous
gas.

The variation of any overall quantity $Q$ can be calculated from the
distribution functions:

 \bea
  \frac{(\frac{\p Q_{tot}}{dt})}{\mathrm{Volume}}&=&\sum_i\int \frac{d^3p_1}{(2\pi)^3
2E_1}\frac{d^3p_2}{(2\pi)^3 2E_2}\frac{d^3p_\ph}{(2\pi)^3
2E_\ph}(2\pi)^4 \de^4(p_1-p_2-p_\ph)|M|^2\nonumber\\
&&[f_2f_\ph (1-f_1)-f_1(1-f_2)(1+f_\ph)] Q_i S_i  \eea

where $S_1=1$ and $S_2=S_\ph=-1$.
\section{Total transverse momentum - with no background}\label{nobackg}

 We want to
calculate the initial transverse momentum when we start with a
standing wave of 1's. The distribution functions are
\bea
f_1&=&\frac{n_1}{2}\left(\de^3(\vec{p_1}-\vec{p_0})+\de^3(\vec{p_1}+\vec{p_0})\right)\nonumber\\
f_2&=&f_\ph=0. \label{eq:dist1} \eea
where $\vec{p_0}$ is the momentum of the standing wave (and $E_0$ will be the corresponding on-shell energy).
Since the two terms initially contribute equally, we are free to
change to one beam instead\footnote{In the numerical calculation we do not change to one beam}

\bea
f_1&=&n_1\de^3(\vec{p_1}-\vec{p_0})\nonumber\\
f_2&=&f_\ph=0. \label{eq:dist2} \eea

This will give the same result.

We calculate the matrix element. From tracing and averaging over
incoming and summing over outgoing spins and assuming the masses of the neutrinos
$m_1=m,m_2=0$:\footnote{Had one chosen a scalar interaction, the $m_1m_2$ term would change sign. However, since we set $m_2=0$ anyway, it makes no difference under this approximation.}

\beq |M|^2=2g^2(p_1\cdot
p_2-m_1m_2)=g^2m^2.\eeq
So we find
\bea
  \frac{(\frac{\p Q_{tot}}{dt})}{\mathrm{Volume}}=-\sum_i\int \frac{d^3p_1}{(2\pi)^3
2E_1}\frac{d^3p_2}{(2\pi)^3 2E_2}\frac{d^3p_\ph}{(2\pi)^3
2E_\ph}(2\pi)^4
\de^4(p_1-p_2-p_\ph)\nonumber\\
g^2m^2n_1\de^3(\vec{p_1}-\vec{p_0}) Q_iS_i\nonumber\\
=\frac{-g^2m^2n_1}{8(2\pi)^5}\sum_i\int
\frac{d^3p_1d^3p_2d^3p_\ph}{E_1E_2E_\ph}
\de^4(p_1-p_2-p_\ph)\de^3(\vec{p_1}-\vec{p_0})Q_iS_i. \nonumber\eea

What we want to find is a measure of the transverse momentum created
in the very beginning. Obviously, there is no momentum if we
just sum over the transverse momentum vectors, so we sum the
magnitudes of created transverse momenta instead. This means putting
$-Q_iS_i=|\vec{p_i}\times \hat{\vec{p_0}}|$. Thus we have

\bea
  \frac{\frac{\p |P_{\bot}|}{dt}}{\mathrm{Volume}}=\frac{g^2m^2n_1}{8(2\pi)^5}\int \frac{d^3p_1d^3p_2d^3p_\ph}{E_1E_2E_\ph}
\de^4(p_1-p_2-p_\ph)\de^3(\vec{p_1}-\vec{p_0})\nonumber\\
\left(|\vec{p_1}\times \hat{\vec{p_0}}|+|\vec{p_2}\times
\hat{\vec{p_0}}|+|\vec{p_\ph}\times \hat{\vec{p_0}}|\right). \eea
After integration we find
\bea
\nonumber \frac{\frac{\p |P_{\bot}|}{dt}}{\mathrm{Volume}}=\frac{g^2m^3n_1}{64(2\pi)^4E_0}. \eea

Inserting $g^2m=\frac{16\pi}{\tau}$ where $\tau$ is the
$1$-lifetime, and $\frac{m}{E_0}=\frac{1}{\ga}$ our final result is

\bea \frac{\frac{d |P_{\bot}|}{dt}}{\mathrm{Volume}}
=\frac{n_1E_0}{8(2\pi)^3\tau \ga^2} \propto \frac{1}{\tau \ga^2},
\eea
as expected.


\section{A more general case}\label{thermback}

In case of a thermal background of neutrinos and pseudoscalars we cannot use
the simple approach specified by Eqs.~\ref{eq:dist1} and \ref{eq:dist2}.
Hence we consider all the distribution functions

\bea
f_2f_\ph (1-f_1)-f_1(1-f_2)(1-f_\ph)
=f_2f_\ph-f_1(1+f_\ph-f_2). \eea

This complicates things a bit. But if we again take only the initial
time it is solvable. The term we are discussing is $|f_2f_\ph-f_1(1+f_\ph-f_2)||\vec{p_i}\times
\hat{\vec{p_0}}|$. We assume initial equilibrium densities of $2$ and $\ph$, which using Boltzmann statistics means that the distribution
functions of $2$ and $\ph$ are identical. It also means that we can use
$f_2f_\ph=\e{-E_1/T}$. Hence we can split the term in an $f_1$ part and an extra
part. The $f_1$ part yields exactly the same result as in Sec.~\ref{nobackg}, but the extra part
is proportional to $f_2f_\ph$ and yields the following:

\bea
 \frac{\frac{\p |P_{\bot}|}{dt}}{\mathrm{Volume}}_{\mathrm{extra}}&=&\int \frac{d^3p_1}{(2\pi)^3
2E_1}\frac{d^3p_2}{(2\pi)^3 2E_2}\frac{d^3p_\ph}{(2\pi)^3
2E_\ph}(2\pi)^4
\de^4(p_1-p_2-p_\ph)\nonumber\\
&&g^2m^2\e{-E_1/T}\left(|\vec{p_1}\times
\hat{\vec{p_0}}|+|\vec{p_2}\times
\hat{\vec{p_0}}|+|\vec{p_\ph}\times \hat{\vec{p_0}}|\right) \eea
which reduces to
\bea \frac{\frac{\p
 |P_{\bot}|}{dt}}{\mathrm{Volume}}_{\mathrm{extra}}=\frac{E_0p_0}{4\pi^2\tau
 \ga}\left(E_0T\e{-m/T}+\int_m ^\infty dE_1\e{-E_1/T} p_1\right) \eea
with relativistic limit
\bea\frac{\frac{\p|P_{\bot}|}{dt}}{\mathrm{Volume}}_{\mathrm{extra}}=\frac{E_0p_0}{4\pi^2\tau
 \ga}\left(E_0T+2T^2\right). \eea

Thus the final result is
\bea
\frac{\frac{\p|P_{\bot}|}{dt}}{\mathrm{Volume}}=\frac{E_0n_1}{8(2\pi)^3\tau
\ga^2}+\frac{E_0p_0}{4\pi^2\tau
\ga}\left(E_0T\e{-m/T}+\int_m ^\infty dE_1\e{-E_1/T} p_1\right) \label{eq:13} \eea

with relativistic limit
\bea \frac{\frac{\p
|P_{\bot}|}{dt}}{\mathrm{Volume}}=\frac{E_0n_1}{8(2\pi)^3\tau
\ga^2}+\frac{E_0p_0}{4\pi^2\tau \ga}\left(E_0T+2T^2\right).
\label{eq:4.18}
\eea

Hence, we are led to conclude that in the realistic scenario of having a background thermal distribution of
light neutrinos as well as of pseudoscalars, we see a correction in the form of the second term in Eq.~\ref{eq:4.18}.
This extra contribution is something which should be taken into account when putting bounds on the neutrino-pseudoscalar interaction.

We can make a rough estimate of the improvement on the bound of the decay coupling constant when taking Eq.~\ref{eq:4.18} into account.
First, we notice that in the presence of the first term on the right-hand side of Eq.~\ref{eq:4.18} we are investigating the standard
case which was previously studied in \cite{Hannestad:2005ex}. Here the naive decay rate $\Gamma_{\rm decay}=\frac{g^2}{16\pi}m$ is translated into a
transport rate $\Gamma_{\rm transport}=\Gamma_{\rm decay}\left(\frac{m}{E}\right)^3$, where the factor $\left(\frac{m}{E}\right)^3$ is due to three Lorentz gamma factors: The first one comes from transforming from the rest frame of the parent neutrino to the frame of the thermal medium. This will give us
a decay rate in the frame of the thermal medium. The other two come from the following reason: The decay is isotropic in the rest frame of the parent neutrino;
however, the decay products will have directions within an angle corresponding to a factor $\gamma$. So, to randomize the direction of the original neutrino
we must include another factor of $\gamma$. In total when we transform from the medium frame decay rate to the relevant transport rate we get two factors of gamma. All in all we arrive at the desired expression
\begin{equation}
 \centering
 \Gamma_{\rm transport}=\Gamma_{\rm decay}\left(\frac{m}{E}\right)^3.
\end{equation}
To ensure that the neutrinos are still free-streaming at the time of photon decoupling as required by observations of the CMB \cite{Hannestad:2004qu,Basboll:2008fx},
we can then compare the transport rate with the expansion rate of the universe $H_{\rm dec}$ at photon decoupling. The requirement for free-streaming is $\Gamma<H_{\rm dec}$.
This leads to the bound \cite{Hannestad:2005ex}
\begin{equation}
 \centering
 g<0.61\times10^{-11}\left(\frac{50\, {\rm meV}}{m_\nu}\right)^2.
\end{equation}
In the event of the decay taking place in a thermal distribution of light neutrinos and pseudoscalars we need to take the second term in Eq.~\ref{eq:4.18} into account.
Especially since at the time of photon decoupling, assuming a generic heavy neutrino mass of $m_\nu=50\, {\rm meV}$ and energy $E=3T_{\nu,\rm{dec}}\sim3\times0.18\,{\rm eV}$, we get $\gamma\sim\frac{E}{m}\sim3.6$.
Combined with the fact that for a relativistic species ($m_\nu<3T_{\nu,\rm{dec}}$) and for our relativistic heavy neutrino $n_\nu\sim T_{\nu,\rm{dec}}^3$ up to factors of order unity,
this means that the transport rate we should be comparing is rather
\begin{equation}
 \centering
 \Gamma_{\rm transport}\sim\Gamma_{\rm decay}\left(\left(\frac{m}{E}\right)^3+16\pi\left(\frac{m}{E}\right)^2\right),
\end{equation}
where the factor of $16\pi$ takes into account this missing factor in the denominator of the second term. If we translate into a bound on the decay coupling constant, this gives
\begin{eqnarray}
 \centering
 g<2.6\times10^{-13}\left(\frac{50\, {\rm meV}}{m_\nu}\right)^{3/2}\left(\frac{T_\nu}{0.18\, {\rm eV}}\right)^{-1/2}
 \left(1+1.8\times10^{-3}\left(\frac{50\, {\rm meV}}{m_\nu}\right)\left(\frac{T_\nu}{0.18\, {\rm eV}}\right)^{-1}\right)^{-1/2}.
\end{eqnarray}
For $\gamma\sim3.6$ with a neutrino mass $m_\nu=50\, {\rm meV}$ this translates simply into the bound
\begin{equation}
 \centering
 g<2.6\times10^{-13},
\end{equation}
i.e. an improvement of more than a factor of 10. Translating this into a limit on the neutrino lifetime in the restframe we get
\begin{equation}
 \centering
 \tau<1\times10^{13}\,\rm{s},
\end{equation}
hence there is still the possibility for the neutrino to be short lived when we let the decay take place in a thermal background.
\section{Numerical results}\label{numerics}
In the numerical implementation we had to change the setup a little.
Because of problems with the bins, it was impossible to get the heavy
neutrino to have vanishing momentum in the transverse direction.
This could not be remedied by increasing the number of bins.
Therefore we made a thermal distribution of $1$'s around an average
momentum. Specifically we chose a standard scenario with
$m=2T,\langle p_x\rangle=\pm3T$ and made a thermal distribution around this. We
chose the artificially high value $g=\frac{1}{\sqrt{40}}$ (making
$T=\frac{320\pi}{\ta}$) to let the code find equilibrium in a
reasonable time. We checked that the code did reach equilibrium both
under the initial condition of no light neutrinos from the
beginning and under the assumption of an initial thermal
distribution of light neutrinos. More on the numerics is provided in the Appendix
\ref{BBHRMOMapp:num}.
\begin{figure}
\centering
\begin{minipage}[c]{0.5\linewidth}
\centering \includegraphics[width=2.5in]{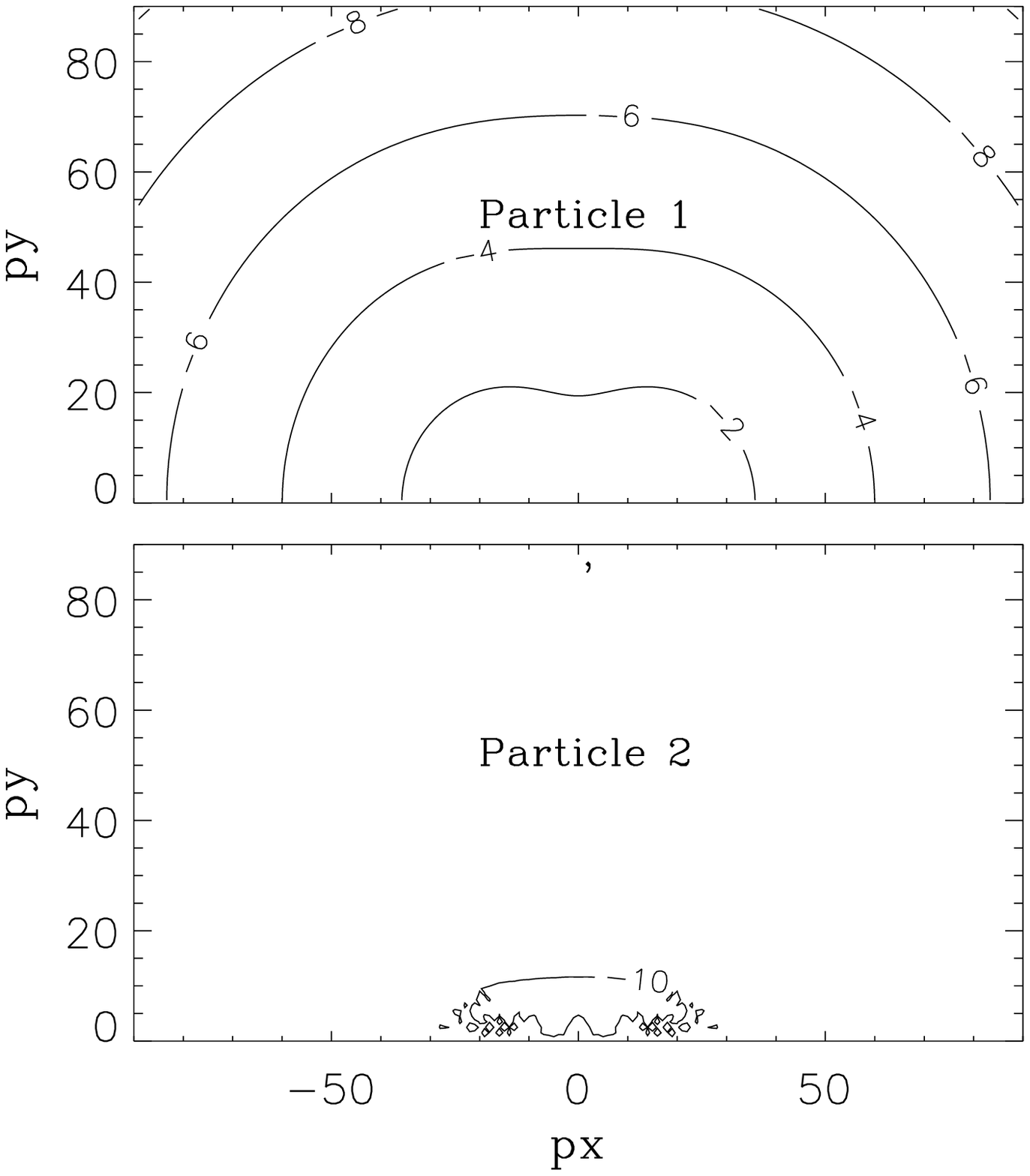} \caption{The
situation just after the first step, $t=9.95\times10^{-9}\,\ta$, in
the standard scenario. The lines are contour plots of the
distribution function $\log(\tilde{f}*T)$ -- as defined in the
Appendix. $p_y$ means transverse momentum and is measured in units
of bin length $dp=\frac{T}{5}$.} \label{BBHRMOMFigStandardstard}
\end{minipage}%
\begin{minipage}[c]{0.5\linewidth}
\centering \includegraphics[width=2.5in]{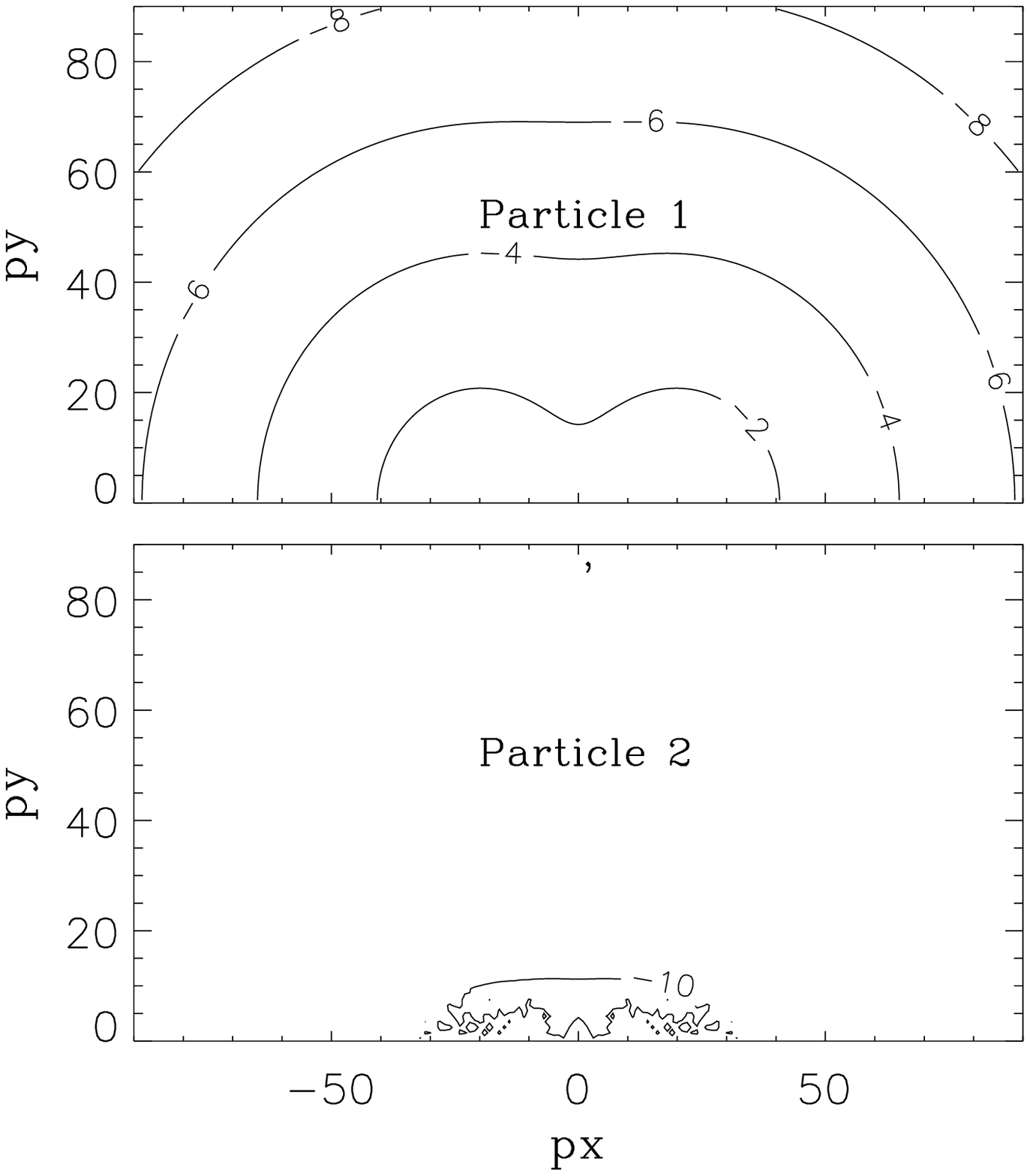} \caption{The
situation just after the first step, $t=9.95\times10^{-9}\,\ta$, in
the alternative scenario. The lines are contour plots of the
distribution function $\log(\tilde{f}*T)$ -- as defined in the
Appendix. $p_y$ means transverse momentum and is measured in units
of bin length $dp=\frac{T}{5}$.} \label{BBHRMOMFigalt}
\end{minipage}
\end{figure}

\begin{figure}
\centering
\begin{minipage}[c]{0.5\linewidth}
\centering \includegraphics[width=2.5in]{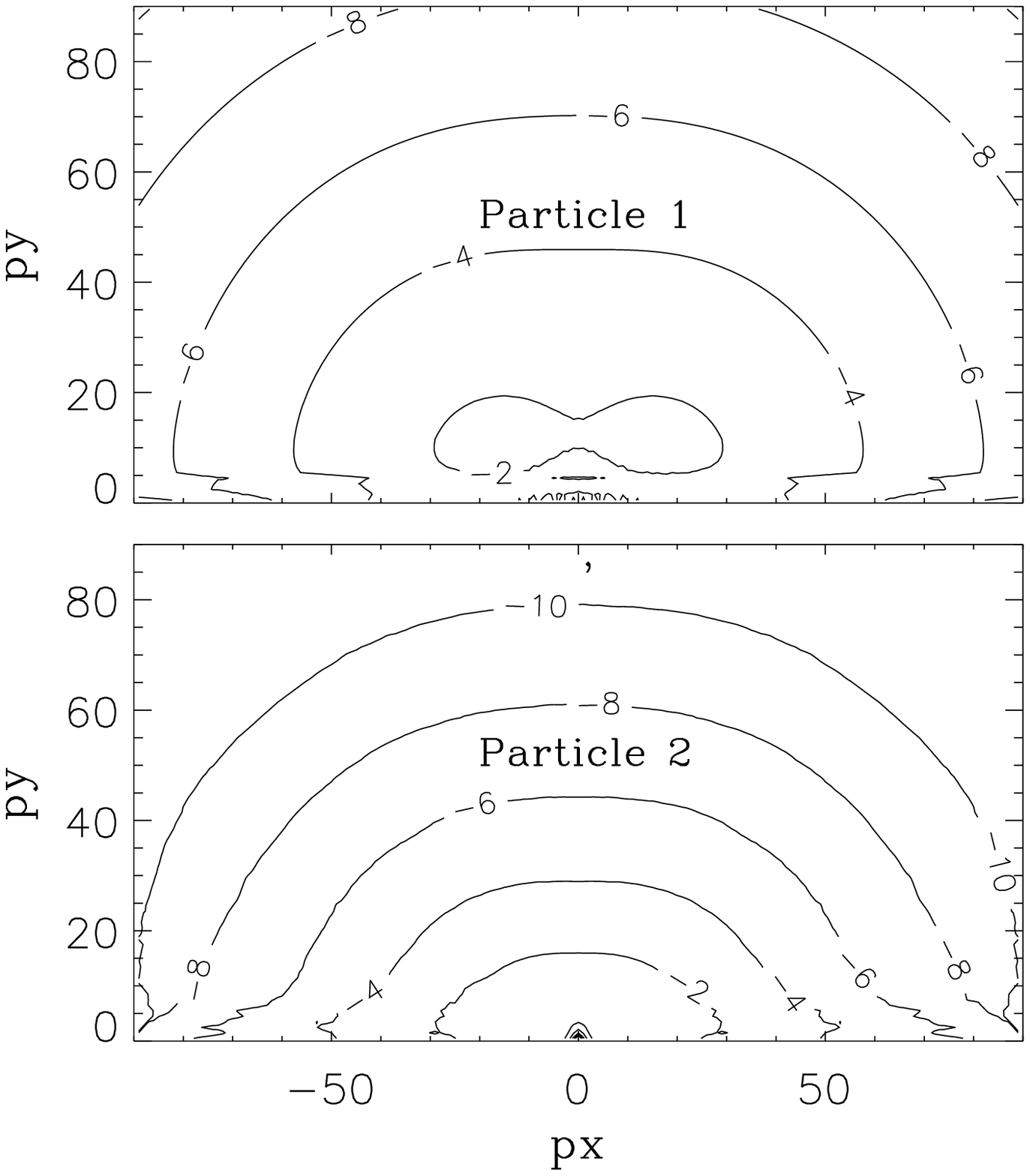} \caption{The
situation later, $t=14.5\,\ta$, in the standard scenario.}
\label{BBHRMOMFigStandardstardlater}
\end{minipage}%
\begin{minipage}[c]{0.5\linewidth}
\centering \includegraphics[width=2.5in]{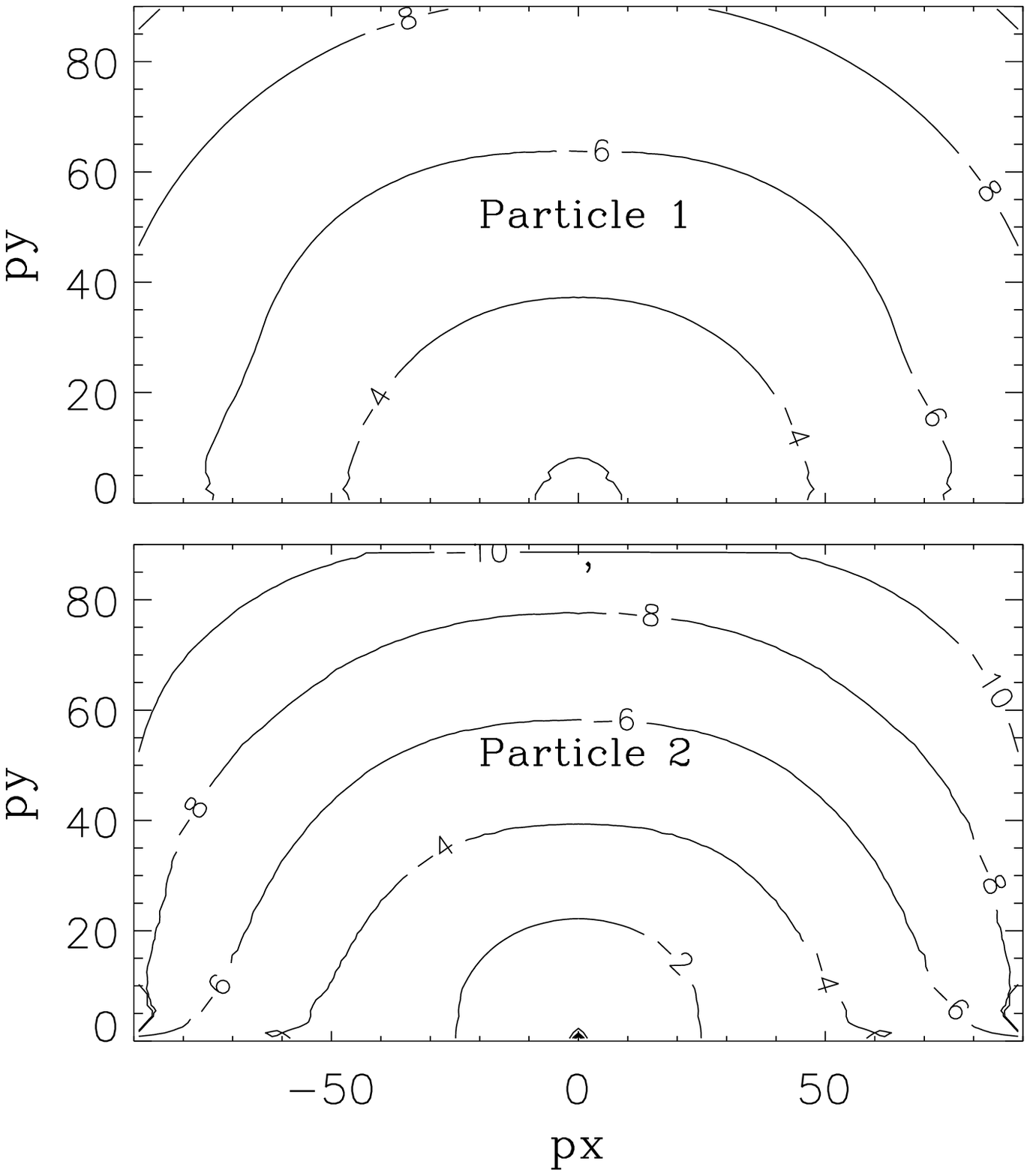}
\caption{The situation even later, $t=1.40\times10^{3}\,\ta$, in the
standard scenario.} \label{BBHRMOMFigStandardstardend}
\end{minipage}
\end{figure}
This means that the calculated formulas cannot be verified
explicitly, since we have no well-defined gamma factor. However, the
factor of $\frac{1}{\tau \ga^2}$ can almost be found. First, the
$\frac{1}{\tau}$ is, of course, trivial. If the coupling is weakened,
the lifetime is correspondingly longer. We checked that our code
yielded this result. The $\frac{1}{\ga}$ for the time dilation of
the neutrino lifetime is also quite trivial. We found this as well -
by noticing the decreased numbers of $2$'s produced in the very
first step when we used an alternative scenario: $m=2T,\langle p_x\rangle=\pm4T$.
Since we had thermal distributions rather than sharply defined
momenta, we could not expect to find the exact relation between
average $\ga\,$s to be the same as the relation between the created
particles.

However, the most interesting second $\frac{1}{\ga}$ can be
illustrated by numerical plots. Fig.~\ref{BBHRMOMFigStandardstard}
shows the first distributions after the very first step, whereas
Fig.~\ref{BBHRMOMFigalt} shows the distributions after the first
step in the alternative scenario with a larger gamma factor. Two
things are very important to notice. First, the created $2$'s are
indeed anisotropic. This is the effect of taking their isotropic
distribution in the frame of the decaying particle $1$ and making a
Lorentz transformation to the (cosmic) laboratory frame. Second,
the fact that the alternative scenario shows more anisotropy among
the $2$'s should make it clear that it must be a gamma factor. One
could alter Eqs.~(\ref{eq:13}), (\ref{eq:4.18}) to match the
initial conditions for the numerics. But since Eqs.~(\ref{eq:13}), (\ref{eq:4.18})
are complicated enough already, and since the numerics is not done to
create new results but only to confirm the pattern of Eqs.~(\ref{eq:13}), (\ref{eq:4.18}),
which it does, we have chosen not to do so.

For completeness, Fig.~\ref{BBHRMOMFigStandardstardlater} shows
the development in the standard scenario at a later time (roughly 15
times the rest-frame lifetime -- this is still an intermediate time
due to the two gamma factors) -- and Fig.~\ref{BBHRMOMFigStandardstardend} shows the standard scenario at an
even later time (roughly 1400 times the rest frame life time), where
equilibrium is almost reached.

One should note that distribution functions are defined according to
the cylinder coordinates used in the code -- see the Appendix
\ref{BBHRMOMapp:num} [especially Eq.~\ref{defftilde} which shows
that the distribution function has dimension of time] for details.
Also, one should note that $p_y$ in the plots are, in fact, the
transverse momentum - not the momentum in one of the transverse
directions. The unit of p in the figures is
$dp=\frac{64\pi}{\ta}=\frac{T}{5}$ which is the distance between
adjacent bins in momentum space.

\section{Conclusion}
We have investigated a neutrino-pseudoscalar gas with only decays  $\nu_1
\to \nu_2+\ph$ and inverse decays to obtain equilibrium. We started
with an anisotropic distribution of $\nu_1$ and confirmed that
kinetic equilibrium is delayed by two Lorentz $\ga$--factors -- one for
time dilation of the heavy neutrino lifetime and one from the
opening angle ie. from the transformation of the isotropic distribution of the
decay products in the rest frame of the decaying particle back to the
(cosmic) laboratory frame. We found explicit analytical expressions
for the rate of creation of transverse momentum -- both in a case
with no background of the decay products and in the case of thermal
backgrounds and the ultrarelativistic limits hereof. We have
confirmed this behavior in numerical simulations as well -- though
we had to make a thermal smear of the initial anisotropy, making the
analytical and numerical results open to a qualitative, but not
quantitative, comparison. 

Furthermore we have obtained updated bounds on the neutrino-pseudoscalar coupling constant
as well as on the neutrino lifetime in the realistic case of a thermal background of
neutrinos and pseudoscalars.

\section*{Acknowledgements}
We would like to thank Steen Hannestad and Georg Raffelt for comments and discussion during the coding and writing part of this project. AB is supported by the Danish Council for Independent Research | Natural Sciences.
\section{Appendix: Numerics} \label{BBHRMOMapp:num}
In order to follow this numerically we notice that we have three
particles in three momentum coordinates - that is, nine dimensions (no
isotropy). We notice that when we assume Maxwell-Boltzmann
statistics particles 2 and $\ph$ behave alike. This means that if
starting conditions are the same, we have to track only one of them.
Even though there is not isotropy, azimuthal angles are arbitrary.
So we end up with four dimensions, two particles, with a momentum in the
initial beam direction and momentum in a transverse direction. So we
want to integrate the remaining coordinates out. We start with

\bea \label{BBHRMOMeq1}C[1]&=&\frac{1}{2E_1}\int
\frac{d^3p_2}{(2\pi)^3 2E_2}\frac{d^3p_\ph}{(2\pi)^3 2E_\ph}(2\pi)^4
\de^4(p_1-p_2-p_\ph)g^2m^2(f_1-f_2f_\ph)\nonumber \\
&=&\frac{dp_{\ph x}dp_{\ph R}dp_{2x}dp_{2R}p_{2R}p_{\ph
R}}{(2\pi)^2E_1E_2E_\ph} \de(E_1-p_2-p_\ph)\de(p_{1x}-p_{2x}-p_{\ph
x})\nonumber\\ &&(f_1-f_2f_\ph)g^2m^2 \int dp_\psi dp_\th
\de(p_y)\de(p_z), \eea

where we have defined (including aligning the coordinate system with
particle 1) momenta \newline
$p_1=(p_{1x},p_{R1},0),p_2=\left(p_{2x},p_{2R}\cos(\psi),p_{2R}\sin(\psi)\right),p_\ph=\left(p_{\ph
x},p_{\ph R}\cos(\th),p_{\ph R}\sin(\th)\right)$.  

After integration we find

\bea C[1]=\frac{dp_{\ph x}dp_{\ph R}dp_{2x}dp_{2R}p_{2R}p_{\ph
R}}{\pi^2E_1E_2E_\ph} \de(E_1-p_2-p_\ph)\de(p_{1x}-p_{2x}-p_{\ph
x})g^2m^2\nonumber\\
(f_1-f_2f_\ph)
\frac{2p_{1R}^2}{\sqrt{S}}\frac{1}{\sqrt{4p_{1R}^2p_{\ph
R}^2-S}+\sqrt{4p_{1R}^2p_{2R}^2-S}}. \eea
where S is given by
\beq \nonumber S\equiv4p_{1R}^2p_{\ph R}^2-\left(p_{2R}^2-p_{\ph
R}^2-p_{1R}^2\right)^2.\label{BBHRMOMdefS}
 \eeq

For numerical purposes, let us underline the formula in the way it
should be implemented.

\bea \frac{df(\vec{p_i})}{dt}=C[i](\vec{p_i}) \eea

however the vectors will not be introduced. Rather we use

\bea \label{defftilde}
 \int_\th f(\vec{p_i})dp_{\th}=\int_\th
 f(p_{ix},p_{iR},p_\th)p_{iR}dp_{\th}
 =2\pi p_{iR}*f(p_{ix},p_{iR},p_\th)\equiv 2\pi
 p_{iR}\tilde{f}(p_{ix},p_{iR})
\eea

and likewise for $C[i]$. This means that the function $\tilde{f}$
that we implement is of dimension $E^{-1}$ and its derivative
dimensionless. The equation implemented is thus

\bea \frac{d\tilde{f}(p_{ix},p_{iR}) }{dt}=\tilde{C}[1]
=\frac{dp_{\ph x}dp_{\ph R}dp_{2x}dp_{2R}p_{2R}p_{\ph
R}}{\pi^3E_1E_2E_\ph}
\de(E_1-p_2-p_\ph)\nonumber\\
\de(p_{1x}-p_{2x}-p_{\ph x})g^2m^2(f_1-f_2f_\ph)
\frac{p_{1R}}{\sqrt{S}}\nonumber\\\frac{1}{\sqrt{4p_{1R}^2p_{\ph
R}^2-S}+\sqrt{4p_{1R}^2p_{2R}^2-S}}, \eea

The implementation of $2$ is quite easy since for fixed momenta $\vec{p_1},\vec{p_2},\vec{p_\ph}$

\bea C[1]=-C[2] \eea or \bea 2\pi p_{1R}\tilde{C}[1]=-2\pi
p_{2R}\tilde{C}[2]. \eea

\section*{References}

\end{document}